# On the initial phenomena occurring in lead/lead collisions at relativistic energies

C. Ythier and G. Mouze

Faculté des Sciences, Université de Nice, 06108 Nice cedex 2, France

mouze@unice.fr

**Abstract**

A new study of the initial phenomena occurring in the fireball should confirm the predicted creation of a new state of nuclear matter having a lifetime of 0.17 yoctosecond (ys) and releasing an energy of 3.87 GeV. The energy-time uncertainty relation might be connected with an up to now unsuspected momentum-position uncertainty relation holding in a three-dimensional time. This new point of view leads to the interpretation of the charge of a particle as being a rotational motion in time, to a new interpretation of inertia, and to a new interpretation of the "color" of a particle. The transverse momentum observed in the study of the fireball might be the signature of this motion in time of the charge.

## 1. Introduction

In a recent communication [1] devoted to the transfer reactions which lead to a complete fusion of projectile and target ,we have shown that the *initial phenomena* occurring in lead/lead, gold/gold or even p/p collisions at relativistic energies should be characterized by the creation of *a new state of nuclear matter*, having a lifetime of only 0.17 ys and causing considerable uncertainties at least in the mass number and the neutron number of the fusion product: This new state has been put into evidence in the fission reaction [2] and in transfer reactions leading to trans-target heavy nuclei [3]. Thus a new study of nucleus-nucleus collisions, realized at moderate beam energies, should confirm these predictions.

But what happens in nucleus/nucleus collisions *if the beam energy is increased* up to relativistic energies? do we have really the *keys* which would be necessary for this new investigation?

One of these keys might be the energy-time uncertainty relation, intimately linked to the most important properties of the new state. The true meaning of this relation has perhaps to be searched for. Indeed, we show in Sect.2 how the *existence of extra-dimensions of time at the border lines of ordinary space* is suggested by this relation.

But arguments in favor of the existence of extra-dimensions of time can be found in a study of the origin of the *rest-mass* of charged particles, or of the origin of *inertia*, or of the meaning of the "*color*" concept, as will be shown in Sect.3.

Sect.4 is devoted to reflections on the symmetry of time and space.

The survival of the *transverse momentum* even in purely head-on collisions might be, as suggested in Sect.5, *the signature of the motion in time* of charged leptons and of "quarks". Does it mean that the fireball has to be studied in the framework of a six-dimensional space-time?

## 2. The meaning of the energy-time uncertainty relation

Since nucleon-phase [2] properties and energy-time uncertainty relation are closely linked, we have first to ask why this important relation, proposed in 1927 by W. Heisenberg [4],

(1) $\Delta E \cdot \Delta t = \hbar$

is nevertheless often considered as "not firmly founded"; and we have further to ask whether new proofs of its validity have not to be searched for.

Indeed, it has been raised as an objection that E and t cannot be defined as operators in the same way as it can be done for $p_x$ and x in the momentum-position uncertainty relation:

(2) $\Delta p_x \cdot \Delta x = \hbar$,

which holds for each direction of the space coordinate [4]. More precisely, it has been raised as an objection that t is only a parameter.

It has been also pointed out that eq.(1) could have been deduced from the observation, made in the study of spectral lines, that the product

(3) (natural line width) X (mean lifetime) is equal to $\hbar$.

Let us show that eq.(1) could have been formulated differently. In high-energy physics, where E becomes much greater than $m_0c^2$ and tends towards pc, it might be tempting to write the first member of eq.(1) as $\Delta p \cdot \Delta t$; but here, by analogy with eq.(2), it might be tempting to give to $\Delta t$ the same dimension as $\Delta x$, in other words to imagine that *t* could be a *three-dimensional time coordinate*; eq.(2) could then be complemented by a *new uncertainty relation*:

(4) $\Delta p_{t1} \cdot \Delta t_1 = \hbar$

 —where $p_t$ is the momentum of the motion in time—,

holding now for each direction of *the time coordinate*.

This apparently absurd hypothesis of the existence of extra-dimensions of time nevertheless deserves serious consideration, as will be demonstrated.

## 3. Three-dimensional time and true nature of the charge

### *3-1 The origin of the rest- mass of a charged particle*

It is well established that "*to any charged particle is attached a rest- mass $m_0$*". Even if the reverse of this sentence is not true --- since to any rest- mass is not attached a charge---, the first expression may be considered as *a principle*.

Let us now consider a charged lepton at rest, for example a negative electron, and let us write the double equation of Einstein

(5) $E = m_0 c^2 = \hbar \omega_0$

This relation shows that "*to any charged particle of mass $m_0$ is associated a given angular frequency $\omega_0$* "

For a negative electron, this angular frequency $\omega_0$ is given by

(6) $\omega_0(e^-) = m_0c^2$ [MeV]$/\hbar$ [MeVs] = 0.510998 MeV / 6.58122 $10^{-22}$ MeVs

   = 7.764 $10^{20}$ rotations /s.

Whereas $\omega_0(\mu^-)$ and $\omega_0(\tau^-)$ are equal to

(7) $\omega_0(\mu^-)$ = 1.60 $10^{23}$ rotations/s,

(8) $\omega_0(\tau^-)$ = 2.70 $10^{24}$ rotations/s.

However, for a particle at rest, *the rotation cannot occur in the ordinary three – dimensional space*, where it is at rest, but only *in time*, and consequently *orthogonally to this 3D-space.*

But a rotation in time can be conceived only in a "three-dimensional" time…

Let us return to eq.(5). Interestingly, the mass $m_0$ of the charged lepton can be written

(5') $m_0 = (\hbar/c^2)\,\omega_0.$

Also $m_0$ depends "exclusively" on $\omega_0$, and we may conclude that *it depends exclusively on an angular frequency of rotation in a 3D- time.*

Let us now return to eq.(4). The *motion* in time – or at least in a region of time having an extent measured by $\Delta t$ --- suggested by this eq.(4) could also *be a rotational motion in a 3D-time.*

Thus any charged lepton, or any quark, could be described by a *chiral flow in time*, orthogonal to the ordinary space.

This description can be extended to the hydrogen atom, as sketched in fig.1 of ref.[5]. In a nucleus, the flow in time of a charge *e* becomes also divided into three components of time, and it is reasonable to admit that *each component carries only one third of the charge.* One sees that *the hypothesis of extradimensions of time* introduces quite naturally the idea of "quarks" carrying either one third or two thirds of the elementary electrical charge – one third being the component of the charge *e* along the $t_3$-axis, and two thirds being the component of the charge in the $(t_1, t_2)$-plane orthogonal to the $t_3$-axis—.

### *3-2 The origin of inertia*

If one assumes that any charge is, in reality, a chiral flow in a three-dimensional time, and that the flow is characterized by an angular frequency $\omega_0$ or a period $T_0$

(9) $\omega_0 = 2\pi / T_0$,

then one must further assume that this rotational motion creates a *centripetal acceleration*

(10) $\gamma' = -\omega_0^2 \, r_0$

or $\gamma' = -4\pi^2 \, r_0 / T_0^2$

*which is orthogonal to the direction of flow.*

Let us now consider a charge of mass $m_0$ in the ordinary three-dimensional space. It is well known that any change, either in magnitude, or in direction, of its velocity requires an acting force, in order to overcome what is known as *inertia*.

But now, we are suspecting that this *obstacle* encountered by an acting force $F = m_0 \gamma$ has been created by the centripetal acceleration described above.

Indeed, the centripetal acceleration certainly tends, *in the 3D-time*, to restrict any change of motion of the charge orthogonally to its direction of flow.

What might be, in the 3D-space, the observable consequences of this situation?

The question deserves to be asked. We first observe that the centripetal acceleration $\gamma'$ *operates* also in the 3D-space, because $\gamma'$ is orthogonal to the direction of flow, which itself is orthogonal to the 3D-space.

Thus, to the tying-down situation holding in the 3D-time must correspond in the 3D-space a similar situation: The charge becomes prevented from moving easily by "inertia", acting in a direction directly opposite to that of its motion, and proportional to the angular frequency $\omega_0$ ( e).

It is the origin of the *inertial mass* of the charge, $m_0 = (\hbar/c^2)\,\omega_0$.

What holds for an isolated charge *e* of mass $m_0$ can be extended to any material body, because any matter is constituted of *charged leptons* and of *quarks*. Thus the force acting on a given material body may be in the 3D-space written

(11) $F = \sum_i n_i\,(\hbar\omega_{0i}/c^2)\cdot\gamma$

where the summation is extended to the various lepton and quark charges, in numbers $n_i$, present in this body. Inertia, this property of any material body, *might also result from the rotational motion in 3D−time of charged leptons and of quarks.*

### 3-3 The true nature of the "color"

In order to abide by the Pauli principle, one multiplies by an antisymmetric "color" wave function the symmetric wave function of particles having a "symmetric quark structure" such as

(12) $|u\,u\,u>$, for the $\Delta^{++}$ particle, or

(13) $|sss>$, for the $\Omega^-$ particle.

However, instead of the antisymmetric color wave function, which uses the three colors red (r), green (g) and blue (b),

(14) $|1> = \dfrac{1}{\sqrt{6}}\,\{|rgb> - |rbg> + |brg> - |bgr> + |gbr> - |grb>\}$,

it would be more satisfactory to use the antisymmetric time wave function, *which uses the three components $t_1$, $t_2$ and $t_3$ of the time coordinate of the charge in the three-dimensional time*:

(15) $|1> = \frac{1}{\sqrt{6}} \{ |t1,t2,t3> - |t1,t3,t2> + |t3,t1,t2> - |t3,t2,t1> + |t2,t3,t1> - |t2,t1,t3> \}$.

The need for a "three−valued charge degree of freedom" expressed in 1964 by O. Greenberg [6] and which led to concept of color, might also have been, in reality, nothing else but the need for extra-dimensions of time.

**3-4 The quantification of the motion in time**

If the uncertainty law expressed by

(4) $\Delta p_{t1} \cdot \Delta t_1 = \hbar$

holds for each direction of the time coordinate, then $p_{t1}$ and $t_1$ have to be considered as "operators".

And "commutation relations", similar to those existing between $p_x$ and $x$, can be established.

Those corresponding to the angular momentum have certainly to play a major role. For example, the angular momentum $\ell_{t3} = t_1 p_{t2} - t_2 p_{t1}$ commutes with $t_3$, but does not commute with $t_1$ or $t_2$.

**4. The point of view of the logician**

The introduction of the concept of a four-dimensional space-time by the theory of relativity was an answer to a need for symmetry.

In a conversation with A. Einstein the great logician K. Gödel is reported to have asked him whether a six-dimensional space-time would not be more symmetrical[1].

The idea, introduced by R.P. Feynman in his theory of positrons [8], that positrons are *electrons moving backwards along the time axis* has constituted a major advance for particle physics, but remained incompatible up to now with a *chiral* flow in time[2]:

---

[1] We are indebted to Prof E. Lohrmann, DESY, for this information
[2] The original idea has nevertheless been translated into his graphs, see [9].

The idea of a chiral motion *in a three-dimensional* time seems to be the logical outcome of Feynman's intuition.

For the e$^-$ -- $\bar{\nu}$ system of the beta-decay, *the rotational motions* in 3D-time of the charged electron and in 3D-space of the neutral antineutrino *should be equally probable*:

Since the angular momentum of this system has necessarily an integer value – the initial and final states having the same mass and the same statistics –, a spin (½) ℏ has to be expected for both the electron and the antineutrino.(This simple reasoning suggests an intuitive picture of the spin).

The classification of leptons and quarks according to three "flavor generations" is nothing else but *a classification according to the value of the angular frequency*: At a given rank, neutrinos and corresponding charged leptons should have *one and the same* angular frequency |$\omega_0$|.

**5. Can the properties of the initial state of the fireball be predicted ?**

This initial state is necessarily the same as that observed in any transfer reaction leading to the complete fusion of projectile and target [1], a state already observed in the reaction of fission [2] and in the formation of trans-target products by ordinary transfer reactions [3].

It is why a new investigation of the fireball could certainly confirm that the state formed at moderate beam energy has the properties of a nucleon phase.

However, even an energy of only 2 AGeV would already be considerably greater than those used at Dubna and at Darmstadt.

It is why we have now to examine the possible consequences of an increase of the beam energy.

*5-1 Consequences of an energy-increase*

The rest-energy of any particle can be converted into binding energy, as it is well known, when this particle combines with another for constituting a new body. But the conversion of the rest-energy of a charged particle or of a quark into kinetic energy is still a quite new question, since the relation giving the increase of mass as a function of velocity still preserves the value of the rest mass $m_0$.

However, according to the views exposed in Sect.3, the rest-energy of a charged lepton or of a quark is essentially an energy of rotation in the three-dimensional time, and this motion is orthogonal to any direction of the three-dimensional ordinary space.

It means that any conversion of the rotation energy in 3D- time might be "observed" as a *momentum* orthogonal to the ordinary space, i.e. at least *orthogonal to the reaction axis* if this change is the result of a nuclear reaction.

It may also be asked *whether the transverse momentum observed in high-energy A/A collisions might be interpreted as the signature of the 3D-time motion of charged leptons and of quarks.*

### *5-2 The true nature of the transverse momentum.*

Indeed, the appearance of a transverse momentum in high energy A/A collisions has been first interpreted as resulting from the existence of an impact parameter [9]. But it was recently reported that this transverse momentum survives even in true "head-on" collisions (see, e.g.,[10]).

Let us consider a simple example, in which the collision occurs between protons.

In his treatise on elementary particle physics [11] C. Berger reports that in the inclusive process occurring in the proton-proton collision

(16) $p + p \rightarrow \pi + X$ ,

−where the final state X can contain several hadrons−

the distribution of the transverse momentum $p_T$ can be represented by the law:

(17) $\dfrac{d\sigma}{dp_T^2} \approx A \exp(-b\, p_T)$

−where $A$ and b are factors which do not depend on $p_T$ and the factor b is equal to about 6 GeV$^{-1}$ −. Berger further reports that this distribution is practically independent from the proton beam energy. Moreover, he reports that the mean value of the transverse momentum furnished by eq.(17) is:

(18)  $\langle p_T \rangle \approx 330$ MeV.

It is noteworthy that this $p_T$-distribution observed below 2 MeV might correspond to "soft" processes, in which the *rest-energy of light charged particles*, such as quarks, *has been totally converted into kinetic energy* in the ordinary 3D-space, but *orthogonally to the reaction axis.*

## 6. Conclusion

A new investigation of the initial step of the nucleus/ nucleus collisions at high beam energy is desirable, because it could not only confirm the formation of the new state of nuclear matter, having a lifetime of 0.17 ys and releasing an energy of up to 3.87 GeV, but also reveal other properties, such as easy nucleon transfer and ephemeral disappearance of the proton charge, and even quite new properties , i.e. properties which up to now were not found in the study of the reaction of fission. With our discussion of the possible meaning of the energy-time uncertainty relation, we hope to have found new keys for a better understanding of what really happens in collisions at relativistic energies.